\documentclass[10pt,conference]{IEEEtran}

\usepackage[utf8]{inputenc}
\usepackage[T1]{fontenc}

\usepackage{flushend}

\usepackage{booktabs}
\usepackage{tabularx}
\usepackage{makecell}

\usepackage{enumitem}
\usepackage{subcaption}

\PassOptionsToPackage{hyphens}{url}\usepackage{hyperref}

\usepackage{tcolorbox}
\tcbuselibrary{theorems}
\usepackage{fontawesome}

\newtcbtheorem{user}{\faWechat{} User Prompt}
{colback=blue!5,colframe=blue!35!black,fonttitle=\bfseries,fontupper=\footnotesize}{th}

\begin{document}

\title{On the Impact of Requirements Smells in Prompts: The Case of Automated Traceability}

\author{
\IEEEauthorblockN{
    Andreas Vogelsang\IEEEauthorrefmark{1}, 
    Alexander Korn\IEEEauthorrefmark{1}, 
    Giovanna Broccia\IEEEauthorrefmark{2}, 
    Alessio Ferrari\IEEEauthorrefmark{3}, 
    Jannik Fischbach\IEEEauthorrefmark{4}, 
    Chetan Arora\IEEEauthorrefmark{5}
}
\IEEEauthorblockA{\IEEEauthorrefmark{1}University of Cologne, Cologne, Germany\\
\{vogelsang,korn\}@cs.uni-koeln.de}
\IEEEauthorblockA{\IEEEauthorrefmark{2}CNR-ISTI, Pisa, Italy\\
giovanna.broccia@isti.cnr.it}
\IEEEauthorblockA{\IEEEauthorrefmark{3} University College Dublin, Dublin, Ireland and CNR-ISTI, Pisa, Italy\\
alessio.ferrari@ucd.ie, alessio.ferrari@isti.cnr.it}
\IEEEauthorblockA{\IEEEauthorrefmark{4}Netlight Consulting GmbH and fortiss GmbH, Munich, Germany\\
jannik.fischbach@netlight.com}
\IEEEauthorblockA{\IEEEauthorrefmark{5}Monash University, Melbourne, Australia\\
chetan.arora@monash.edu}
}

\maketitle

\begin{abstract}
 Large language models (LLMs) are increasingly used to generate software artifacts, such as source code, tests, and trace links. Requirements play a central role in shaping the input prompts that guide LLMs, as they are often used as part of the prompts to synthesize the artifacts. However, the impact of requirements formulation on LLM performance remains unclear. In this paper, we investigate the role of requirements smells---indicators of potential issues like ambiguity and inconsistency---when used in prompts for LLMs. We conducted experiments using two LLMs focusing on automated trace link generation between requirements and code. Our results show mixed outcomes: while requirements smells had a small but significant effect when predicting whether a requirement was implemented in a piece of code (i.e., a trace link exists), no significant effect was observed when tracing the requirements with the associated lines of code. These findings suggest that requirements smells can affect LLM performance in certain SE tasks but may not uniformly impact all tasks. We highlight the need for further research to understand these nuances and propose future work toward developing guidelines for mitigating the negative effects of requirements smells in AI-driven SE processes.
\end{abstract}


\section{Introduction}
In the rapidly evolving field of AI, prompt engineering has become a key area of interest, focusing on developing and optimizing prompts for large language models (LLMs). Developers are increasingly relying on LLMs for a variety of software engineering (SE) tasks, such as generating code from requirements~\cite{Liu2022}, deriving test cases from requirements~\cite{fischbach23,Arora24}, or tracing requirements to code~\cite{rodriguez2023prompts}. These tasks often hinge on the quality of the requirements used to prompt the LLM, yet the impact of the specific formulation of requirements on LLM performance is still poorly understood.

Our new idea is to examine the potential role of \textit{requirements smells}---indicators of issues like ambiguity or inconsistency in natural language (NL) requirements---in prompt effectiveness. In SE, requirements smells have long been recognized as potential obstacles that can lead to misinterpretations and lower quality in the software development process~\cite{Femmer17}. If smells are present in the requirements used as input for LLMs, they could similarly impact the quality of the LLM's outputs in SE tasks.

We conducted experiments using two leading generative LLMs, GPT-4o and Llama 3.1, to explore this phenomenon in the context of automated trace link generation between requirements and code. This task is particularly relevant, as LLMs are increasingly used to automate traceability---a crucial process that ensures alignment between high-level requirements and their implementation in code~\cite{lin2021traceability,rodriguez2023prompts}. We compared the LLMs' performance on trace link generation using 94 requirements and associated code (809 LOC) from five projects (70 trace links), considering non-smelly and smelly requirements.

Our emerging results present a mixed picture. While requirements smells had a small but statistically significant negative impact on predicting if a requirement was implemented in a code segment, they had no significant effect on identifying the specific lines of implementation. This suggests that LLMs can 
cope with low-quality requirements for the investigated task and the limited complexity of the systems we tested. However, there are also indications that smells affect certain aspects of LLM performance but their impact varies by task and smell type.


These preliminary findings open several research avenues, such as exploring the relevance of smells in other requirements-centric SE tasks, understanding the influence of project scale and domain, developing methods to identify and mitigate smells, and investigating \textit{prompt smells}---the broader implications of ``smell'' in prompt engineering~\cite{Ronanki24}.

\section{Background and Related Work}



\textbf{Requirements Smells:}
``Requirements smells'' refer to patterns or characteristics in software requirements that indicate potential issues, leading to problems in downstream development activities~\cite{Femmer17}. These smells can signify ambiguous, incomplete, inconsistent, or overly complex requirements, resulting in increased costs, delays, or defects in the final product~\cite{Femmer2019}.
Frattini et al.~\cite{Frattini2022} published a catalog of 206 requirements quality indicators (aka \textit{smells}) extracted from a systematic mapping study on 105 relevant primary studies~\cite{Montgomery2022}. The authors also categorized requirements smells into three categories: (1)~\textit{lexical} smells describe issues in single words or terms, e.g., code = ``program source'' or ``set of rules''?; (2)~\textit{syntactic} smells describe issues in word or sentence structures, e.g., \textit{When the system sends a message to the receiver, it shall provide an acknowledgment} (it = ``system'' or ``receiver''?); and (3)~\textit{semantic} smells describe issues in interpreting the requirements within its context, e.g., \textit{The system shall generate a report at the end of each day}, (strictly at midnight or at the end of the business hours?). We sampled three types of requirements smells from each of the three categories of this catalog and used them in our study. 

\textbf{Automated Traceability:}
The ability to trace relevant software artifacts to support reasoning about the quality of the software and its development process plays a crucial role in requirements and software engineering, particularly for safety-critical systems~\cite{Guo24}.
LLMs have led to considerably better results in solving trace link recovery (TLR) as a classification task~\cite{lin2021traceability}. Rodriguez~et~al.~\cite{rodriguez2023prompts} tested the capability of Claude, a generative AI model, to solve TLR tasks directly by prompting it with NL instructions. 
North~et~al.~\cite{North2024} use explainable AI techniques to trace input requirements to LLM generated code. 

Our paper is the first to analyze the effect of requirements smells on trace link generation. Additionally, our roadmap extends beyond traceability, exploring the impact of smells on other generative SE tasks.


\section{Study Design}

We aim to assess the impact of requirements smells in prompts on downstream SE tasks. We analyze automated traceability as a representative task, considering the following research questions (RQs):

\begin{description}
    \item [\textbf{RQ1:}] \textbf{How well can LLMs trace high-quality requirements to existing code?}
    \item [\textbf{RQ2:}] \textbf{How does the presence of requirements smells in prompts impact the tracing performance?}
    \item [\textbf{RQ3:}] \textbf{How do different requirement smell categories impact the tracing performance?}
\end{description}




To answer the RQs, we perform a benchmark study on five manually curated projects. We evaluate the tracing performance of LLama 3.1 and GPT-4o on well-written requirements (RQ1) and in smelly ones, considering the impact of the number of smells (RQ2) and the smell category, i.e., lexical, syntactic, semantic (RQ3).

\subsection{Study Objects and Data Collection}
We experiment on five exemplary cases with 94 requirements of which 70 were implemented in 809 LOC. All five cases represent simple games, which are more or less popular. One of the authors created a list of requirements for all five cases and a Java code file that implements a fraction of the requirements. 
Another author has reviewed the list of requirements and the code to ensure correctness and sufficient quality. 
We also created a smelly version for a subset of the requirements. Each smelly requirement represents exactly one type of smell. We sampled nine smell types from a recently published catalog of requirements smells (see Table~\ref{tab:smells})~\cite{Frattini2022}. 
Table~\ref{tab:studyObjects} gives an overview of the considered cases including their number of requirements, how many requirements are implemented in the code, how many requirements have a smelly variant, and how large the code is. 


\begin{table}
\scriptsize
    \centering
    \caption{Study Objects}
    \label{tab:studyObjects}
    \begin{tabular}{@{}lrrrr@{}}
    \toprule
     \textbf{Game} & \textbf{\#req.} & \textbf{\#implemented req.} & \textbf{\#req. with smelly variant} & \textbf{\#LOC}\\
    \midrule
     Dice & 25  & 19 & 18 &141 \\
     Arkanoid & 19 & 14 & 15 &152\\
     Snake & 14 & 7 & 11 &142 \\
     Scopa & 16 & 15 & 13 &220\\
     Pong & 20 & 15 & 15 &154\\
     \midrule
     Sum & 94 & 70 & 72 & 809 \\
    \bottomrule
    \end{tabular}
    \vspace{-2em}
\end{table}

\begin{table*}
\scriptsize
    \centering
    \caption{Number and types of studied smells (categorization based on Frattini et al.~\cite{Frattini2022})}
    \label{tab:smells}
    \begin{tabularx}{\textwidth}{@{}lXrrrrrr@{}}
    \toprule
     \textbf{Smell categories and types}& \textbf{Description}&\textbf{Dice} & \textbf{Arkanoid} & \textbf{Snake} & \textbf{Scopa} & \textbf{Pong} & \textbf{Sum}\\
    \midrule
    \textbf{Lexical smells} & &\textbf{7}& \textbf{4}& \textbf{4}& \textbf{4}& \textbf{5}& \textbf{24}\\
\hspace{1em} subjective language & Words of which the semantics are not objectively defined, such as \textit{user-friendly}, \textit{easy to use}, \textit{cost effective}, etc. \newline
Sentences expressing personal opinions or feelings.
& 5& 2& 2& 2& 3& 14\\
\hspace{1em} optional parts &Sentences containing optional parts, e.g., by using the words \textit{possibly}, \textit{eventually}, \textit{if possible}, \textit{if needed}, etc.& 1& 1& 1& 0& 2& 5\\
\hspace{1em} weak verbs &Weak verbs, such as \textit{can}, \textit{could}, \textit{may}, etc.& 1& 1& 1& 2& 0& 5\\ [1ex]
    \textbf{Syntactic smells} && \textbf{6}& \textbf{4}& \textbf{3}& \textbf{3}& \textbf{4}& \textbf{20} \\
\hspace{1em} vague pronouns &Pronouns that refer back to a previous part of the text for which the reference is unclear.& 1& 1& 1& 1& 1& 5\\
\hspace{1em} passive voice &Sentences using passive voice such that it is unclear who is performing a certain action.& 3& 2& 1& 2& 2& 10\\
\hspace{1em} negative phrases &Sentences containing negative modifiers (e.g., \textit{not}) or negative expressions.& 2& 1& 1& 0& 1& 5\\ [1ex]
    \textbf{Semantic smells} && \textbf{5}& \textbf{7}& \textbf{4}& \textbf{6}& \textbf{6}& \textbf{28}\\
\hspace{1em} logical inconsistencies& Two requirements, which are connected to the same concepts, contradicting each other.& 2& 3& 1& 1& 1& 8\\
\hspace{1em} numerical discrepancies & Two requirements connected to the same concepts, containing inconsistent and/or contradicting numerical information. & 1& 1& 1& 1& 1& 5\\
\hspace{1em} ambiguities & Unclear/imprecise sentence parts that can be misunderstood if read by different people.& 2& 3& 2& 4& 4& 15\\
\midrule
\textbf{Sum} & & \textbf{18}&\textbf{15}&\textbf{11}&\textbf{13}&\textbf{15}&\textbf{72}\\
    \bottomrule
    \end{tabularx}
    \vspace{-1em}
\end{table*}

\textbf{Ground truth creation:} 
Two authors created a ground truth by independently reviewing the smell-free requirements and deciding which requirements are implemented in the code in which LOC ($\kappa = 0.9$). Similarly, two authors reviewed the smelly requirements and decided which should be considered as implemented in the code ($\kappa= 0.74$). Any remaining differences were resolved in two meetings.


\textbf{Ratio of Smelly Requirements}:
We investigate the effect of diverging levels of requirements quality by assessing the tracing performance. We define the level of requirements quality by the ratio of smelly requirements.

\textbf{Smell Category}:
In RQ3, we are interested in the impact of different smell categories. We test the differences between three smell categories: lexical, syntactic, and semantic smells. Each category is represented by three smell types associated with the category (see Table~\ref{tab:smells}).

\textbf{Language Model}:
We ran our experiments on two generative LLMs, GPT-4o (gpt-4o-2024-08-06) and Llama 3.1 (70B). Both models have a 128k token context window. We selected these models as they represent the most advanced closed- and open-source LLMs at the time of the study.

\subsection{Tracing Performance Measures}
We measure tracing performance by examining an LLM's ability to predict whether a requirement is implemented and in which LOC.

\textbf{Binary tracing:}
We define binary tracing accuracy (BTA) as the accuracy of an LLM's prediction on the question of whether or not a requirement is implemented in the code:
\begin{equation*}
    \mathit{BTA} = \frac{|\mathit{implemented} \wedge \mathit{traced}| + |\neg\mathit{implemented} \wedge \neg\mathit{traced}|}{N}
\end{equation*}

\noindent where $N$ is the total number of considered requirements.

\textbf{LOC tracing:}
We assess the ability to trace a requirement to locations in the code by measuring the precision, recall, and $F_1$ of the traced LOC compared with the true lines of code implementing a requirement.
\begin{equation*}
\textit{LOC precision} = \frac{|\textit{LOC is traced} \wedge \textit{LOC implements req}|}{|\textit{LOC is traced}|}
\end{equation*}
\begin{equation*}
\textit{LOC recall} = \frac{|\textit{LOC is traced} \wedge \textit{LOC implements req}|}{|\textit{LOC implements req}|}
\end{equation*}

$F_1$ is the harmonic mean between precision and recall. LOC precision, recall, and $F_1$ are computed for each requirement. To answer our RQs, we compute the mean LOC precision, recall, and $F_1$\footnote{Mean $F_1$ may lie outside the range of mean precision and recall due to averaging differences and the harmonic mean’s sensitivity to extremes.} of all requirements implemented in the code; we neglect the LOC tracing for requirements not implemented. 




\subsection{Study Execution}

We crafted a prompt by following the experience with automated traceability prompts reported by Rodriguez~et~al.~\cite{rodriguez2023prompts}. When manually creating the ground truth, we created and followed specific tracing instructions, which we also included in the prompt. The prompt is included in our online material\footnotemark{}.
After executing the prompt, the LLM outputs a JSON object containing the requirement ID, a yes/no field indicating whether the model thinks the requirement is implemented, and a list of numbers indicating the LOCs the model associates the requirements with. 

\footnotetext{\url{https://doi.org/10.6084/m9.figshare.27153441}}



To control any non-deterministic behavior by the LLMs, we set the temperature to zero~\cite{Peng2023}. We executed each prompt five times and used majority voting to determine the final answer.
For a game with $n$ requirements and $m$ smelly versions available, we replaced $p$ of the requirements with their smelly counterparts in each run. We did $2 + 120$ runs, one for each extreme case ($p=0$ and $p=m$), and 120 for the intermediate cases, with a random selection of $p$. The quality level of each run was represented by the percentage of smelly requirements in the total set $(p/n)$. The 122 runs allowed us to cover different quality levels while keeping the workload manageable.
In total, we ran 3,050 prompts on each LLM, covering five games with 122 samples per game and five runs (majority voting) per sample.

\section{Study Results and Discussion}
\subsection{RQ1: General Tracing Performance}

Table~\ref{tab:rq1_results} shows the binary tracing accuracy (BTA) and the LOC tracing precision, recall, and $\text{F}_1$-score achieved in the runs with 0\% smelly requirements. 
The average BTA is 0.95 for GPT-4o and 0.96 for Llama 3.1, indicating high reliability of the LLMs for this task when considering relatively small, yet realistic, projects like the one included in our study. For a more complex task such as LOC tracing, the performance decreases, but remains within acceptable boundaries, with GPT-4o outperforming Llama 3.1 by 0.08 in terms of $\text{F}_1$, making it the preferred model for this task.

\begin{table}

\scriptsize
    \centering
    \caption{Tracing performance with 0\% smelly reqs.}
    \label{tab:rq1_results}
    \begin{tabular}{@{}lrrrrrrrr@{}}
    \toprule
     & \multicolumn{4}{c}{\textbf{GPT-4o}} & \multicolumn{4}{c}{\textbf{Llama 3.1}} \\  
    \cmidrule(lr){2-5}\cmidrule(l){6-9}
    \textbf{Game} & \textbf{BTA} & \makecell{\textbf{LOC} \\ \textbf{prec.}} & \makecell{\textbf{LOC} \\ \textbf{rec.}}& \textbf{$\text{F}_1$}& \textbf{BTA} & \makecell{\textbf{LOC} \\ \textbf{prec.}} & \makecell{\textbf{LOC} \\ \textbf{rec.}}& \textbf{$\text{F}_1$}\\
    \midrule
dice & 0.96 & 0.71 & 0.83 & 0.73 & 0.92 & 0.68 & 0.73 & 0.65 \\
arkanoid & 1.00 & 0.68 & 0.80 & 0.69 & 0.95 & 0.64 & 0.61 & 0.53 \\
snake & 0.93 & 0.61 & 0.75 & 0.63 & 0.93 & 0.49 & 0.67 & 0.53 \\
scopa & 1.00 & 0.74 & 0.84 & 0.74 & 1.00 & 0.64 & 0.70 & 0.63 \\
pong & 0.90 & 0.59 & 0.73 & 0.59 & 0.95 & 0.65 & 0.71 & 0.64 \\
\midrule
Mean & 0.96 & 0.67 & 0.79 & 0.68 & 0.95 & 0.62 & 0.68 & 0.60 \\
    \bottomrule
    \end{tabular}
    \vspace{-2em}
\end{table}

\begin{figure*}
\vspace{-0.2cm}
    \centering
    \begin{subfigure}{0.32\textwidth}
        \centering
        \includegraphics[width=\textwidth]{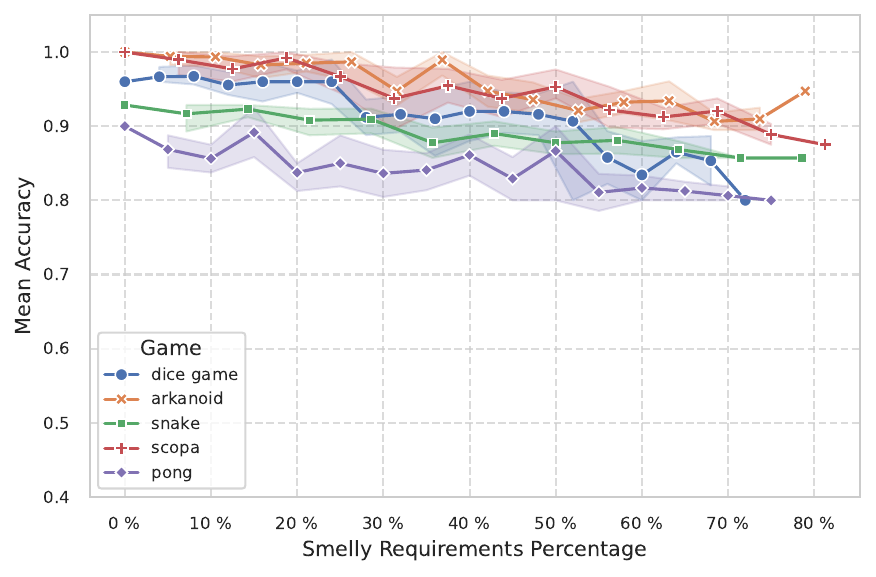} 
        \caption{Binary tracing accuracy (GPT 4o)}
        \label{fig:rq2_results_bta}
    \end{subfigure}
    \begin{subfigure}{0.32\textwidth}
        \centering
        \includegraphics[width=\textwidth]{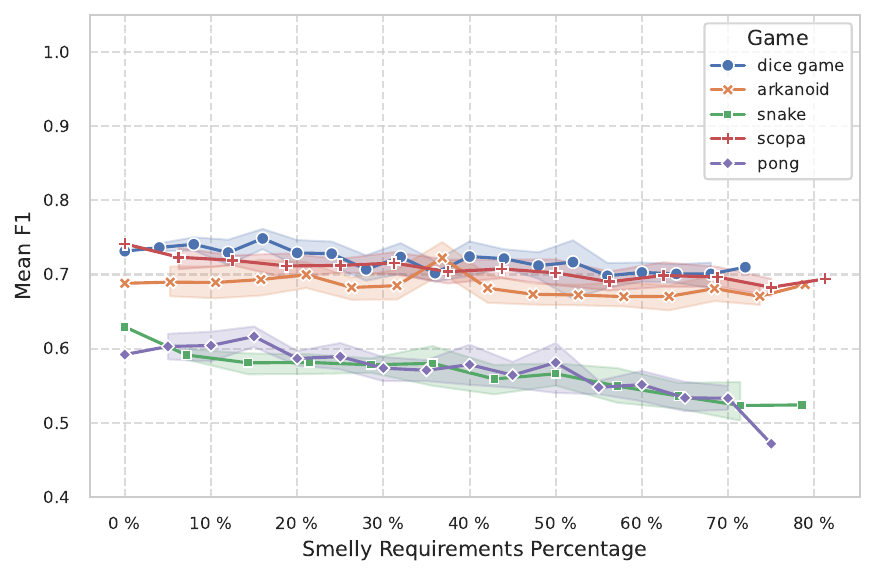} 
        \caption{LOC tracing $\text{F}_1$-score (GPT 4o)}
        \label{fig:rq2_results_LOCf1}
    \end{subfigure}
    \vskip \baselineskip
    \vspace{-1em}
    \begin{subfigure}{0.32\textwidth}
        \centering
        \includegraphics[width=\textwidth]{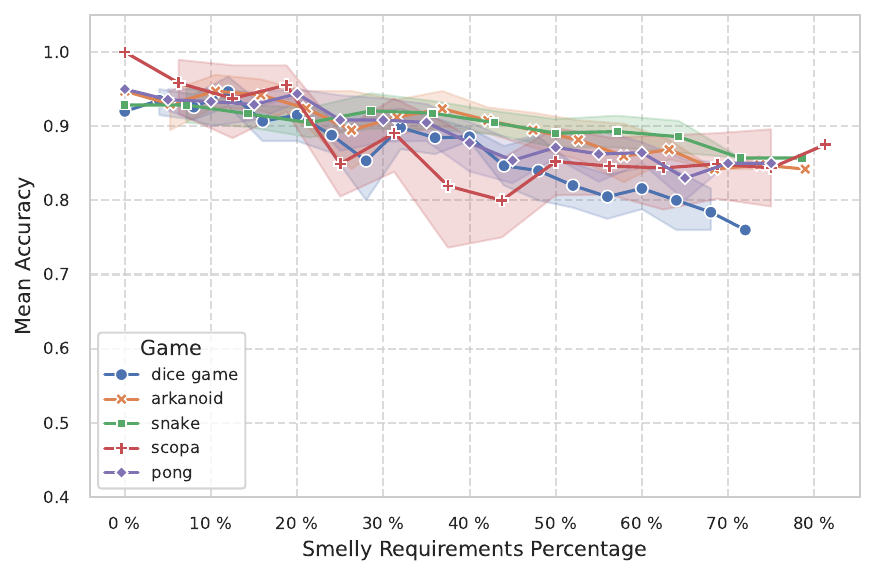} 
        \caption{Binary tracing accuracy (Llama 3.1)}
        \label{fig:rq2_results_bta_llama}
    \end{subfigure}
    \begin{subfigure}{0.32\textwidth}
        \centering
        \includegraphics[width=\textwidth]{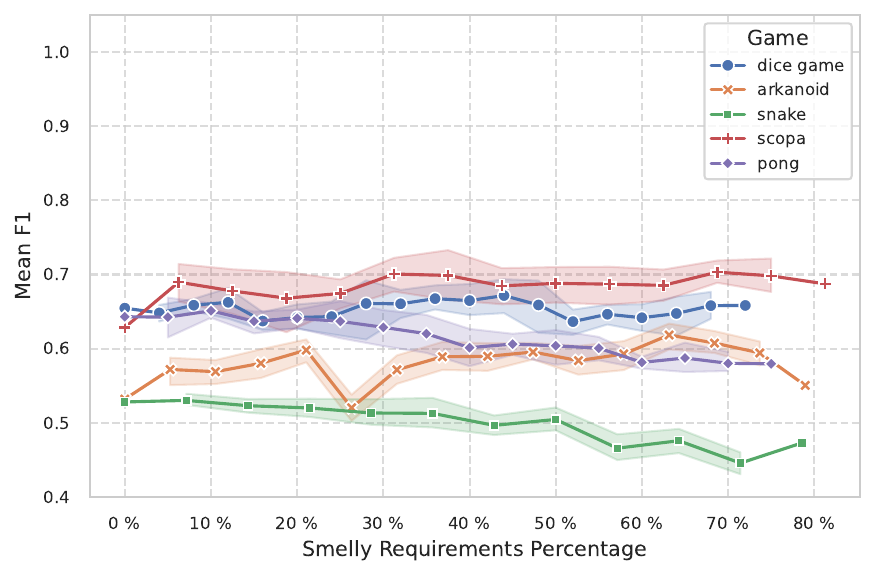} 
        \caption{LOC tracing $\text{F}_1$-score (Llama 3.1)}
        \label{fig:rq2_results_LOCf1_llama}
    \end{subfigure}
    \caption{Results for RQ2: Tracing performance with increasing ratio of smelly requirements.}
    \label{fig:rq2_results}
    \vspace{-0.4cm}
\end{figure*}

\subsection{RQ2: Tracing Performance with Smells}
Fig.~\ref{fig:rq2_results} shows the BTA and the LOC tracing $F_1$ score achieved in the runs with increasing ratios of smelly requirements. To quantify the relation between the ratio of smelly requirements and the performance, we have fitted a generalized linear mixed model (GLMM) on our data with \textit{game} and \textit{LLM} as random effects. This means the model is fitted to predict the performance (BTA, $F_1$) based on the percentage of smelly requirements and conditioned on the corresponding game and the used language model. The model for BTA converges and shows a good fit. A summary of the model can be found in our online material\footnotemark[\value{footnote}]. The model coefficient is -0.001, meaning that for every 10\% increase in smelly requirements, the BTA decreases by 0.01. A p-value~<~0.001 indicates that this effect is statistically significant. The confidence interval $[-0.002, -0.001]$ suggests that this negative impact is consistent across the games and language models.
The model for $F_1$ converged but the model coefficient of $-0.001$ was not statistically significant ($p=0.055$). 

\vspace{-0.2cm}
\subsection{RQ3: Tracing Performance for Smell Categories}

Table~\ref{tab:rq3_results} shows the mean BTA and the mean LOC tracing precision, recall, and $\text{F}_1$-score achieved for the smelly requirements over all samples of our experiments grouped by the three categories.
While syntactic smells (e.g., vague pronouns, passive voice, negative phrases) seem less problematic, the tracing performance for requirements with semantic smells (e.g., inconsistencies, ambiguities) was generally worse. This implies that one should prioritize strategies to avoid these types of smells when using LLMs for trace link generation.   

\begin{table}
\scriptsize
    \centering
    \caption{Tracing performance for smell categories}
    \label{tab:rq3_results}
    \begin{tabular}{@{}lrrrrrrrr@{}}
    \toprule
     & \multicolumn{4}{c}{\textbf{GPT-4o}} & \multicolumn{4}{c}{\textbf{Llama 3.1}} \\  
    \cmidrule(lr){2-5}\cmidrule(l){6-9}
    \makecell{\textbf{Smell} \\ \textbf{category}} & \textbf{BTA} & \makecell{\textbf{LOC} \\ \textbf{prec.}} & \makecell{\textbf{LOC} \\ \textbf{rec.}}& \textbf{$\text{F}_1$}& \textbf{BTA} & \makecell{\textbf{LOC} \\ \textbf{prec.}} & \makecell{\textbf{LOC} \\ \textbf{rec.}}& \textbf{$\text{F}_1$}\\
    \midrule
lexical & 0.90 & 0.69 & 0.76& 0.69 & 0.83 & 0.65 & 0.68 & 0.69\\
syntactic & 0.98 & 0.74 & 0.84& 0.73 & 0.91 & 0.76 & 0.75 & 0.73\\
semantic & 0.83 & 0.63 & 0.76& 0.63 & 0.86 & 0.64 & 0.70 & 0.63\\
    \bottomrule
    \end{tabular}
\vspace{-2em}
\end{table}

\subsection{Discussion and Conclusion}
We conclude that current LLMs show good performance in the analyzed tracing tasks on high-quality requirements. Moreover, they can also cope with low-quality requirements for the investigated task and the limited complexity of the systems we tested.
On the other hand, the significant effect of smelly requirements on BTA and the differences in smell types and games suggest that investigating the effect on larger systems and other SE tasks is worthwhile.



\vspace{-0.1cm}
\section{Future Plans} 

We aim to study the impact of requirements smells in prompts on software development processes where LLMs are used to generate artifacts. To understand this relationship, we will explore several key research avenues.

\textbf{Effect of Smells in Requirements-Centric Tasks:} We plan to adapt our experiments to other SE tasks, such as code generation, model synthesis, and test case derivation, to examine if and how requirements smells affect these activities. We will assess the correlation between requirements smells and generated artifact smells, such as code or test smells~\cite{di2018detecting,spadini2018relation}. In addition to quantitative analysis, we will gather qualitative insights to identify emerging problems in generated artifacts. Expert opinions will complement performance results to ensure LLM outputs are measured and explained, following the approach of Ferrari et al.~\cite{ferrari2024}.

\textbf{Effect of Project Scale and Domain:} We plan to replicate our experiments on larger, more complex systems. We hypothesize that the effect of smells on downstream tasks will be stronger in complex systems due to the increased difficulty LLMs face handling complex contextual information~\cite{shi2023large}. Additionally, domain specificity may affect LLM performance, especially if the model lacks domain-specific training~\cite{gururangan2020don}.

\textbf{Smell Identification and Correction:} To mitigate the impact of smells, prompt issues must be addressed. We will investigate self- and human-assisted correction techniques, such as self-correction by LLMs or clarification queries from the analyst. This will guide the development of practical LLM-based solutions for requirements review, scaling beyond individual requirements to application-level scenarios~\cite{mu2024clarifygpt}.


\textbf{Beyond Requirements Smells:} 
Prompts express requirements~\cite{Vogelsang2024} and, like requirements, may suffer from quality issues. It is unclear how linguistic quality affects LLM performance or how \textit{requirement smells} translate to \textit{prompt smells}~\cite{Ronanki24}. We aim to assess whether ‘smelly’ prompts impact benchmark tasks (e.g., math reasoning, counterfactual evaluation) or if new \textit{prompt smells} are needed.




\bibliographystyle{IEEEtran}
\bibliography{references}

\end{document}